\documentclass[twocolumn]{article}

\usepackage[pdftex, hidelinks, pdftitle={High-Frequency Radar Ocean Current Mapping at Rapid Scale with Autoregressive Modeling}]{hyperref}

\usepackage[pdftex]{graphicx}
\usepackage{tikz}
\usepackage{pgfplots}
\usepackage{xcolor}
\pdfoutput=1

\usepackage{amsmath}
\usepackage{amssymb}
\usepackage{gensymb}

\newcommand\vect[1]{\boldsymbol{#1}}
\renewcommand{\mathbb}[1]{\mathbf{#1}}
\everymath{\displaystyle}

\usepackage{multirow}
\usepackage{array}

\begin{document}

\title{High-Frequency Radar Ocean Current Mapping at Rapid Scale with Autoregressive Modeling\thanks{This work has been accepted for publication in IEEE Journal of Ocean Engineering.}}
\author{Baptiste Domps, Dylan Dumas, Charles-Antoine Gu\'erin, Julien Marmain \thanks{B. Domps and J. Marmain are with the Radar \& Scientific Applications Department, Degreane Horizon, Cuers, France (e-mail: \{baptiste.domps,julien.marmain\}@degreane-horizon.fr). D. Dumas and C.-A. Gu\'erin are with the Universit\'e de Toulon, Aix-Marseille Univ., CNRS, IRD, MIO UM 110, Mediterranean Institute of Oceanography, Toulon, France (e-mail: \{dylan.dumas,guerin\}@univ-tln.fr).}}


\maketitle


\begin{abstract}
We use an Autoregressive (AR) approach combined with a Maximum Entropy Method (MEM) to estimate radial surface currents from coastal High-Frequency Radar (HFR) complex voltage time series. The performances of this combined AR-MEM model are investigated with synthetic HFR data and compared with the classical Doppler spectrum approach. It is shown that AR-MEM drastically improves the quality and the rate of success of the surface current estimation for short integration time. To confirm these numerical results, the same analysis is conducted with an experimental data set acquired with a 16.3 MHz HFR in Toulon. It is found that the AR-MEM technique is able to provide high-quality and high-coverage maps of surface currents even with very short integration time (about 1 minute) where the classical spectral approach can only fulfill the quality tests on a sparse coverage. Further useful application of the technique is found in the tracking of surface current at high-temporal resolution. Rapid variations of the surface current at the time scale of the minute are unveiled and shown consistent with a $f^{-5/3}$ decay of turbulent spectra.
\end{abstract}



\section{Introduction}

Ocean surface current mapping with High-Frequency Radars (HFR) is now a well-established technique and is used routinely for a growing number of civil, environmental and scientific applications (e.g. \cite{roarty2019,heron2016,paduan2013high}). The dominant Bragg diffraction mechanism for scattering of HF/VHF electromagnetic waves from the sea surface was first unveiled in the fifties by Crombie \cite{crombie1955}. It took 2 more decades before the complete physical theory to describe sea clutter and its application to ocean current extraction was achieved \cite{crombie1972,barrick1972,barrick1974,stewart1974,barrick1977}. While the dramatic progress in instrumentation, computerization and costs reduction explain in large part the success and expansion of HFR systems in the last years, there have been no major breakthrough in the main physical picture nor in the signal processing techniques whose progress have been incremental.

Virtually all radar systems softwares still rely on a spectral analysis of the backscattered time series to infer the value of the radial surface current from the frequency shift of the dominant Bragg ray in the sea Doppler spectrum \cite{gurgel1999}. This is sufficient for the vast majority of applications, where the integration time is large enough (typically, from 10~min to 1~h) to ensure both adequate frequency resolution and satisfactory Signal-to-Noise ratio (SNR) of the first-order Bragg peaks.

However, for some emerging applications such as tsunami early warning \cite{dzvonkovskaya2018,guerin_OD18}, ship tracking \cite{laws2016} or prediction of Lagrangian transport in oil spill \cite{berta2014}, short integration times are required, a constraint which deteriorates the accuracy and reliability of surface current estimation and reduces the radar coverage. There has been some isolated attempts in the literature to propose alternative, non-spectral, signal processing methods to improve the quality of oceanic parameters \cite{guerin2018}. In particular, the use of Autoregressive (AR) modeling was applied by \cite{khan1991} within the framework of iceberg detection in HFR coastal measurements. Later on, \cite{martin1997} first proposed an AR approach in the context of oceanographic HFR, whose interest has been confirmed through subsequent studies \cite{middleditch2006,wang2015,wang2017}. To date, however, these methods have been confined to specific applications and have not yet earned their letters of nobility in the common HFR signal processing toolbox. 

In this paper, we revisit the AR modeling and apply it to systematic surface current mapping. We show that an AR model combined with a Maximum Entropy Method (MEM, \cite{johnparkerburg1975}) for the optimal estimation of the AR model coefficients makes it possible to drastically enhance the magnitude of the first-order Bragg peaks in the sea clutter, as compared to conventional Fourier analysis. When this technique is combined with classical quality criteria it yields a significant improvement of both the reliability and the coverage of estimated radial current for short integration times (of the order of 1~minute).


The paper is organized as follows. Section \ref{sec:autoregressive} describes the mathematical framework used to process the oceanic backscattered signal with an AR model and the MEM technique for the calculation of the coefficients. Section \ref{sec:performance} assesses the performances of the model with synthetic HFR data and compares them with a classical Fourier analysis. Section \ref{sec:application} presents a first application of this methodology to an actual HFR data set recently acquired in the Mediterranean sea in the region of Toulon for sea surface currents mapping. The validity of this novel approach to produce short-time estimation of surface current is assessed in Section \ref{sec:drifters} through some experimental comparisons with surface drifters data. Section \ref{sec:time} investigates the rapid time variations of sea surface current at fixed location; it is shown that the AR-MEM method paves the way to high temporal-resolution monitoring of radial currents. A notable result is the experimental evidence of a turbulent behavior with a typical $-5/3$ spectral decay for small-scale process down to the minute time scale.

\section{Autoregressive Model for the Oceanic Doppler Spectrum}\label{sec:autoregressive}

\subsection{Oceanic Doppler Spectrum}\label{sec:oceanic_doppler_spectrum}

The sea surface radar cross section (RCS) per unit bandwidth per unit area $\sigma(\omega)$ was theoretically expressed by \cite{barrick1972,barrick1972_RS} as a functional perturbation expansion with respect to a small dimensionless parameter, which can be seen as the ratio of wave height to radar wavelength. This expansion is limited to second-order, which is sufficient to capture the main observed features:
\begin{equation}\label{defsigma}
  \sigma(\omega)=\sigma_1(\omega)+\sigma_2(\omega)
  \end{equation}
The dominant, first-order term is limited to a symmetric pair of spectral rays at the so-called Bragg resonant frequency:
\begin{equation}
    \label{eq:1st_spectrum_nospeed}
    \sigma_1(\omega)=2^6\pi K_0^4\Big\{S_d(-2\vect{K_0})\delta(\omega-\omega_B)+S_d(+2\vect{K_0})\delta(\omega+\omega_B)\Big\}
\end{equation}
with
\begin{equation}\label{omegab}
\omega_B=2\pi f_B=2\pi\sqrt{\frac{g}{\pi\lambda}}
\end{equation}
Here $g=9.81$ m.s$^{-2}$ is the gravity constant, $\lambda$ is the radar wavelength, $S_d$ is the directional wave density energy spectrum, $\vect{K_0}$ is the incident horizontal wave vector and $\delta$ the Dirac function. The second-order term $\sigma_2(\omega)$ is a continuous, regular function of frequency obtained from a quadratic summations of the wave spectrum. It contains the proper contribution of non-resonant waves (including wind wave and swell) to the scattering process. Its complicated expression needs not be detailed here and can be found in many references \cite{barrick1972_RS,lipa1986}.

In presence of a surface current of (positive or negative) radial speed $U_r$ the first- and second-orders terms are uniformly shifted by an extra frequency:
\begin{equation}\label{omegac}
    \omega_c=\frac{4\pi U_r}{\lambda}
\end{equation}
resulting in an overall translation of the Doppler spectrum:
\begin{equation}
    \label{eq:1st_spectrum_speed}
    \sigma(\omega)=\sigma_1(\omega-\omega_c)+\sigma_2(\omega-\omega_c)
\end{equation}
For bistatic radars such as the one used here (see \cite{dumas_OD2020} for a detailed description), eqs. \ref{omegab} and \ref{omegac} must be adapted to the system geometry by introducing the bistatic angle $\varphi$, being half the angle between transmitter, range cell and receiver:
\begin{equation}
	\omega_B=2\pi\sqrt{\frac{g}{\pi\lambda}\cos\varphi}~~\textrm{and}~~\omega_c=\frac{4\pi U_r}{\lambda}\cos\varphi
\end{equation}

As it is well known, the estimation of surface current from HFR data is based on the identification of the first-order Bragg rays in the oceanic Doppler spectrum and the calculation of their extra shift $\omega_c$ with respect to the theoretical Bragg frequency. This process is rendered difficult not only by the presence of noise (which limits the estimation range) but also by the second-order contribution which can induce parasitic peaks in the vicinity of the Bragg frequency. Hence, when assessing the theoretical performances of any surface current estimation procedure, it is important to model the full second-order Doppler spectrum.

\begin{figure}[h]
    \centering
    \includegraphics[width=\linewidth]{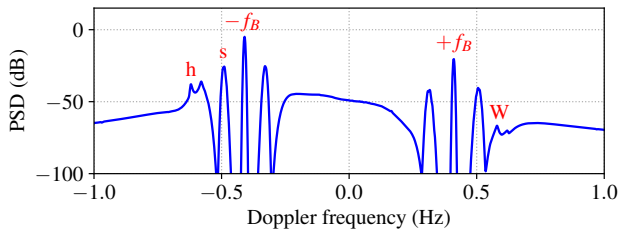}
    \caption{Simulated bistatic ($\varphi=15\degree$) second-order Doppler spectrum at radar frequency 16.3~MHz, for a Pierson-Moskowitz oceanic spectrum with wind speed $5$~m.s$^{-1}$ in the absence of surface current and noise. The first-order Bragg peaks are located at $\pm f_B$. Second-order peaks are marked by ``s'' (swell peaks); ``W'' (wind waves peaks); ``h''  (hydrodynamical peaks) \cite{grosdidier2013}.}
    \label{fig:spectresamuel}
\end{figure}

In order to obtain a realistic synthetic radar signal, a numerical simulator for second-order bistatic Doppler spectra was used and tested in \cite{grosdidier2013} (Figure \ref{fig:spectresamuel}). A statistical reference Doppler spectrum, $\sigma_0$, can thus be generated under typical sea states conditions for the Mediterranean sea in the absence of noise and currents. An example of such spectrum for a Pierson-Moskowitz oceanic spectrum by a $U_{10}=5$~m.s$^{-1}$ wind speed is shown in Figure \ref{fig:spectresamuel}. As seen, the second-order components of the Doppler spectrum exhibit peaks of different origins (swell, wind wave local maxima, hydrodynamical kernels) which can cause false detection when proceeding to a systematic research of the main Bragg rays.

\subsection{Classical Spectral Estimator}

As it is well-known, the theoretical Doppler spectrum (eq. \ref{defsigma}) is the limiting form of the Power Spectral Density (PSD) of the backscattered time series $s(t)$ for large integration time and surface patch (to within an amplitude factor given by the radar equation):

\begin{equation}\label{psd}
    \sigma(\omega)\sim\left\langle\left\vert\int_0^T e^{-i\omega t}s(t)W(t)\,dt\right\vert^2\right\rangle
\end{equation}

Here $W(t)$ is a windowing function over the integration time $T$ and the complex radar signal $s(t)$ is assumed to be resolved in range and direction. Note that such azimuthally resolved Doppler spectra can only be obtained with phased-arrays using a beam forming process. In the case of compact antenna arrays using Direction Finding techniques, the available Doppler spectrum can only be described by an azimuthal integration of eq. \ref{eq:1st_spectrum_nospeed}.

In most radar systems, the range-processed ``I'' and ``Q'' voltage signals are usually acquired at a constant temporal rate $\Delta t$ equal to the chirp duration providing uninterrupted sequences of size $N$. Calculation of the Fourier Transform (eq. \ref{psd}) is achieved through Fast Fourier Transform (FFT) of the resulting complex times series $s(n\Delta t)$ with, typically, a Blackman-Harris tapering window. This leads to an estimation of the Doppler spectrum with $N$ spectral rays at the frequency resolution $\Delta f=1/T$. In the following, we will henceforth simply refer to this estimation technique as the ``FFT method''. Both the accuracy and reliability of the radial current estimation are limited by the integration time. The resolution to which the current can be determined from the spectral rays of the radar signal even in the absence of noise  is in principle limited by the frequency resolution ($\Delta U_r = \lambda/2T$). The actual accuracy can be to some extent improved by the centroid method which allows choosing intermediate frequencies between the spectral rays \cite{barrick1980}. The reliability of the estimation in presence of noise (that is, the ability to detect a ``true'' Bragg line) depends primarily on the SNR \cite{forget_RS15}, which can be shown proportional to the integration time. Hence using short integration times deteriorates the quality of the estimation in a twofold aspect. As we will show, the use of AR modeling makes it possible to mitigate simultaneously the issues of weak SNR and poor frequency resolution.

\subsection{Autoregressive Representation of Time Series}

The aforementioned limitations of the Fourier Transform in calculating Doppler spectra are not specific to HFR and actually arise in a wealth of applications. Several alternative approaches have been proposed since the late seventies \cite{stoica2005}. One family of methods relies on a stochastic representation of times series with help of AR modeling and has become popular in certain fields such as statistics and econometrics. It has been first shown \cite{marple1977} that AR can achieve better spectral resolution than FFT on a canonical, two-sinus case. Later on, \cite{schlindwein1992} demonstrated the ability of AR models to successfully represent short-time ultrasonic Doppler signals in the case of blood flow estimation, which are formally similar to sea surface backscattered signals under the first-order Bragg approximation. Rather than using a Fourier series, AR assumes that the signal $s(n\Delta t)$ at any present time step $n$ can be written as a weighted sum of its previous values in the past, together with a white noise $\varepsilon[n]$:
\begin{equation}
    \label{AR}
    s(n\Delta t)=-\sum_{k=1}^pa[k]s\big((n-k)\Delta t\big)+\varepsilon[n]
\end{equation}
The number $p$ of involved time steps in the past is called the order of the AR representation and the $a[k]$ are termed ``AR coefficients''. The choice of the order and the computation of the AR coefficients will be discussed further. The PSD of the AR process (eq. \ref{AR}) is given by (e.g. \cite{marple2019}):
\begin{equation}
    \label{eq:psd_ar}
    P_{AR}(\omega)=P_\varepsilon {\left|1+\displaystyle\sum_{k=1}^pa[k]e^{-\textrm{i}\omega k\Delta t}\right|}^{-2}
\end{equation}
where $P_\varepsilon$ is the constant PSD of the white noise $\varepsilon$. Note that the calculation of the PSD requires no further assumption on the signal besides its AR representation (eq. \ref{AR}). Contrarily to the Fourier Transform approach, it does not suffer from truncation effect such as Gibbs phenomena and does not requires tapering windows. The expression of the AR PSD (eq. \ref{eq:psd_ar}) is mathematically equivalent to the PSD calculated with a FFT with the major difference that it can be evaluated at arbitrary frequencies $\omega$ (not only integer multiples of the lowest frequency) once the AR coefficients $a[k]$ are known. This is very advantageous for short samples where the discrete frequency set prescribed by FFT is limited by a coarse resolution. Nevertheless, the AR modeling is also subject to hidden limitations for small order $p$ and short sample size $N$. A better representation of the PSD requires higher orders while an accurate evaluation of the AR coefficients (most often based on the signal auto-correlation function) requires large sample size ($N\gg p$). As we will see in the following, the best trade-off is obtained for $p\sim N/2$.

\subsection{Maximum Entropy Spectral Estimation}

Efficiency of the AR spectral calculation does not only lay in eq. \ref{eq:psd_ar} but primarily in the method chosen for obtaining the AR coefficients $\{a[k]\}_{k=1}^p$. Hence, inaccurate estimations of $\{a[k]\}_{k=1}^p$ will result in a poor spectral representation. Many methods were developed over the past half century for the determination of these coefficients, e.g. least squares, likelihood, Yule-Walker and Burg. We refer e.g. to \cite{marple2019} for a description and comparison of these methods. As a general rule, each AR model is equivalent to a linear prediction filter, that is to say, every sample value $s(n\Delta)$ can be predicted as a linear weighted sum of the previous $p$ sample values (eq. \ref{AR}). The Burg's method uses an error criterion to recursively compute the $n$th AR coefficient $a[n]$ from $a[n-1]$ by minimizing both the forward and backward prediction errors. The main asset of the Burg's algorithm is its well-known ability to provide good estimations when applied to short signals \cite{laeri1990}, which is the reason why we selected this method in the present context. This method is nowadays widely available in most signal processing toolboxes. Minimizing the total prediction error energy is equivalent to a maximum information entropy estimation. In agreement with the radar literature, spectral estimation based on Burg's method will therefore be entitled as Maximum Entropy Method (MEM) in the following.

\subsection{Accuracy criterion for radial current estimation}\label{sec:criterion}

The estimation of radial surface currents from the backscattered Doppler spectrum is usually deteriorated by various factors such as poor SNR, Radio Frequency Interferences (RFI), beam forming artifacts due to insufficient rejection of secondary lobes, rapid variation of the bistatic angle leading to varying Bragg frequency within the radar cell or widening the Bragg peaks due to spatial variations of the current in the farthest radar cells. Dubious values can be eliminated by a series of quality checks (QC, e.g \cite{gomez2014}), the simplest of which rely on a SNR threshold for the Bragg rays (typically SNR~$>$~6~dB), an \textit{a priori} limitation of the search interval (typically $|U_r|<$~1~m.s$^{-1}$) and consistent Doppler shifts for the positive ($f^+$) and negative ($f^-$) Bragg peaks which must undergo the same translation, that is to say $\vert f^+-f^--2f_B\vert$ should be small. Application of successive QC reduces increasingly the available area in the surface current map. When using short integration time in the FFT method, the QC are hardly met and dramatically reduces the coverage. The main advantage of the AR-MEM approach is to increase the SNR and to provide a better filling of surface current map with the same constraint of accuracy. In the analysis of the experimental data set we imposed the following stringent QC:
\begin{equation}\label{QC}
  \begin{split}
    \textrm{SNR}(f^\pm)& > 12~\textrm{dB}\\
    \vert f^+-f^--2f_B \vert&<0.025 f_B
    \end{split}
\end{equation}
The accuracy threshold at 16.15~MHz corresponds to a current speed discrepancy of 0.1~m.s$^{-1}$ and the choice of 12~dB results from an empirical trade-off between coverage and reliability.

\section{Performance assessment on synthetic data}\label{sec:performance}

Because of the claimed performances of the MEM in the literature, a significant gain in spatial coverage and accuracy is expected for the current maps. In order to quantify these increased performances, we went through a series of tests using synthetic HFR time series with a known radial surface current. In the absence of noise, the time variation of the received complex voltage signal over a time interval $[0,T]$ and acquired at rate $\Delta t$ can be described by a random stationary process with prescribed Doppler spectrum:
\begin{equation}\label{s0}
    s_0(t) = \sqrt{\Delta t}\sum_{j=1}^N\sqrt{\sigma_0(\omega_j)}e^{\textrm{i}(\omega_j+\omega_c)t}e^{\textrm{i}\varphi_j}
\end{equation}
where $\omega_j=2\pi j/T$, the $\varphi_j$ are uniform random phases and as before $\omega_c$ is the frequency shift (eq. \ref{omegac}) due to the surface current and $\sigma_0(\omega)$ is the simulated second-order sea surface Doppler spectrum in the absence of surface current and noise (see Figure \ref{fig:spectresamuel}). The noise $n(t)$ is modeled as an additive random process with adjustable level so that the received signal $s(t)$ can be written:
\begin{equation}
    \label{eq:signal_bruite}
    s(t) = s_0(t) + \alpha n(t),
\end{equation}
where $\alpha>0$ is a dimensionless coefficient controlling the relative level of noise (hence the SNR). Here, both the signal $s_0$ and noise $n$ are normalized to have unit variance. It is customary to model the noise process $n(t)$ with a complex Gaussian white noise. However, it turns out that the actual noise characteristics inferred from the radar data deviate from the normal distribution. Therefore, we also devised an hybrid synthetic model, where the numerical signal $s_0(t)$ generated from eq. \ref{s0} is combined with an experimental additive noise $n(t)$. To obtain the actual noise component from real measurements we used the so-called ``RFI'' files provided by the WERA software. These files record the negative frequencies in the FFT range-gating operation and only contain the noise part of the signal (as they correspond to ``negative'' time-delays). They can be used to obtain realistic samples for the noise.

A key parameter for accurately representing the time series with an AR model is the order $p$ (eq. \ref{AR}). As seen from eq. \ref{eq:psd_ar}, it governs the spectral resolution of the PSD. Choosing higher orders will thus result in a better frequency representation of the modeled time series. However, as the input time series is of finite duration, these higher orders will also produce edge effects because of the lag in the AR linear prediction filter (eq. \ref{AR}) and eventually deteriorate the resulting modeled spectrum.

\begin{figure}[h]
    \centering
    \includegraphics[width=\linewidth]{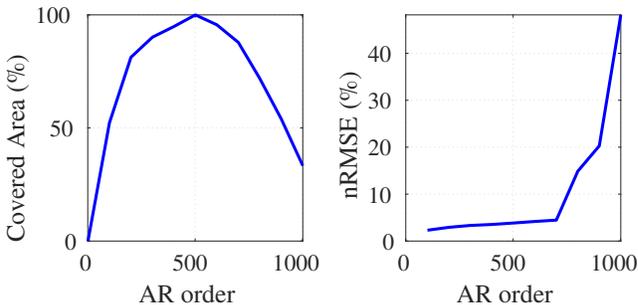}
    \caption{Influence of the AR order in estimating the radial current $U_r$ from a synthetic HFR time series with AR-MEM for an integration time of 4.5~min ($N=1024$~time steps) and $U_r=0.3$~m.s$^{-1}$. The radar signal is modeled by a random stationary process with prescribed Doppler spectrum (eqs. \ref{s0} and \ref{eq:signal_bruite}) and a Gaussian  white noise with $\alpha=1$ ($\textrm{SNR}=0~\textrm{dB}$). The left plot shows the proportion of Doppler spectra (in \%, among 1000 samples) passing the quality test as a function of the AR order; the right panel shows the normalized error (nRMSE, in \% of $U_r$) in estimating the radial current.}
    \label{fig:ordre_ar}
\end{figure}

To determine the optimal order, we performed a series of numerical tests according to the model (eqs. \ref{s0}, \ref{eq:signal_bruite}) for some typical values of the radial current $U_r$ and for a wide range of sample sizes $N$ and AR order $p$. The radar frequency (16.15~MHz) and the sampling rate ($\Delta t=0.26$~s) were taken identical to the actual HFR system in Toulon. The Doppler spectrum $\sigma_0(\omega)$ was simulated according to the second-order model (eq. \ref{defsigma}) with a Pierson-Moskowtiz spectrum by wind speed $U_{10}=5$ m.s$^{-1}$. The noise was chosen to be a Gaussian white noise with $\alpha=1$ (i.e. $\textrm{SNR}=0~\textrm{dB}$). We went through a systematic application of the AR-MEM to estimate the radial current from the first-order Bragg peaks following the QC (eq. \ref{QC}). For each test case, a large number (1000) of random time series were produced and the proportion of samples passing the quality test was calculated together with the Root-Mean-Square Error (RMSE) between estimated and actual radial current. Figure \ref{fig:ordre_ar} shows an example of such simulation for a typical sample of size $N=1024$ (about 4.5~min) and radial current $U_r=0.3$ m.s$^{-1}$. As seen, an optimal success rate is obtained for an order about $p=500$ while the RMSE remains almost constant for orders smaller than about $p=700$ and deteriorates rapidly for higher orders. This observed behavior reflects the trade-off that must be found between high orders and edge effects in the calculation of the prediction filter. The following empirical rule was retained after thorough numerical studies: the optimum order $p$ is half of the number $N$ of samples in the input time series (Figure \ref{fig:ordre_ar}). In practice, choosing orders smaller than $N/2$ does not deteriorate much the SNR of the AR PSD nor the RMSE in the estimation of radial current but reduces significantly the computational time.

\begin{figure}[h]
    \centering
    \includegraphics[width=\linewidth]{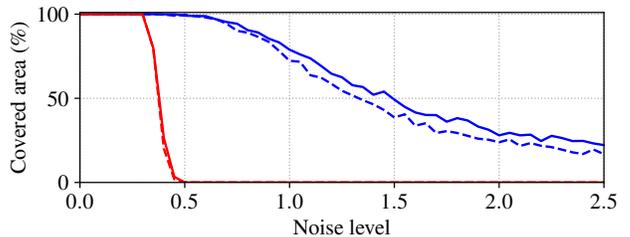}
    \caption{QC success rate ($\%$) of the Doppler spectra for short ($N=1024$ time steps) synthetic HFR times series, when computed with {\bfseries\color{blue} ------}~AR-MEM and experimental noise; {\bfseries\color{red} ------}~FFT method and experimental noise; {\bfseries\color{blue} -\,-\,-\,-}~AR-MEM and Gaussian white noise; {\bfseries\color{red} -\,-\,-\,-}~FFT method and  Gaussian white noise.}
    \label{fig:theo_comparaisons}
\end{figure}

Once the optimal AR order has been identified we have compared the respective performances of the FFT and AR-MEM methods for short samples, in particular the robustness to noise. Synthetic random data sets with the same signal characteristics as above but with various noise level ($0\leq \alpha\leq 2$, i.e. $-3~\textrm{dB}<\textrm{SNR}<0$~dB) were generated for the two types of noise (white Gaussian noise or experimental noise). The performance of each method have been quantified in terms of rate of success in passing the QC test (Figure \ref{fig:theo_comparaisons}) over 1000 samples for each noise level. A drastic gain in robustness with respect to the two types of noise is observed when using the AR-MEM instead of the FFT method as the maximum noise level allowing for a full rate of success is twice larger.

\section{Application to HFR surface current mapping}\label{sec:application}

\begin{figure}[h]
    \centering
    \includegraphics[width=\linewidth]{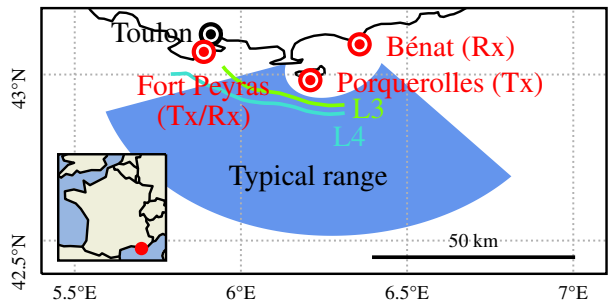}
    \caption{Radar sweep area of the multistatic setup of Porquerolles (transmitting site), Cap B\'enat (receiving site) and Fort Peyras (not used here) along with drifter's trajectories on June 24-25th, 2019 ({\bfseries\color{green} ------} $L_3$, {\bfseries\color{cyan} ------} $L_4$).}
    \label{fig:intro_carto}
\end{figure}

The University of Toulon and the Mediterranean Institute of Oceanography (MIO) are operating a multistatic HFR network (Figure \ref{fig:intro_carto}) in the region of Toulon \cite{guerin2019, dumas_OD2020}. It is composed of a transmitting site on Porquerolles Island, a receiving site at Cap B\'enat and a transmitting/receiving site at Fort Peyras (see Figure \ref{fig:intro_carto}). The working radar frequency is 16.15~MHz with a frequency band of 100~kHz, leading to a range resolution of 1.5~km. The receiving sites in Fort Peyras and Cap B\'enat were originally equipped with arrays of 8 antennas which could not allow for the application of beam forming techniques for the azimuthal discrimination of surface currents and could only be processed with direction finding techniques. Recently, the receive sites were upgraded to linear arrays of 12 antennas (Cap B\'enat as of January 2019 and Fort Peyras as of January 2020), which now opens the possibility to apply beamforming techniques and therefore allows the use of azimuthally resolved voltage time series. In this section, we present an analysis of such time series arising from the Cap B\'enat site in the light of the AR-MEM. The data set used as a benchmark for the derivation of radial surface current maps was recorded on February 4, 2019, 14:01~UTC.

\begin{figure}[h]
	\centering
	\includegraphics[width=\linewidth]{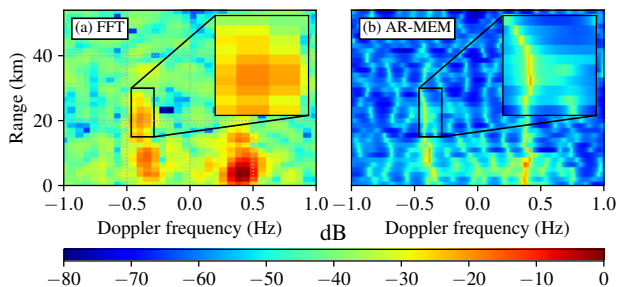}
    \caption{Range-Doppler maps (dB; colorscale) computed with short samples ($N=128$ time steps, $T=33$~s) along the central azimuth for the data acquired on Feb. 4th, 2019, by the Cap B\'enat WERA HF radar at 14:01~UTC. The PSD has been sampled using the: (a)~FFT ($\Delta f = 2,3 \cdot 10^{-2}~\textrm{Hz}$); (b)~AR-MEM with optimal order ($p=64$) and 4096 frequency steps ($\Delta f = 9,4 \cdot 10^{-4}~\textrm{Hz}$). Both insets are centered on the negative Bragg line ($-f_B$) and rescaled to a $\pm 1~\textrm{m}.\textrm{s}^{-1}$ window.}
	\label{fig:compare_psd}
\end{figure}

We first evaluated the performances of AR-MEM in computing the PSD by comparing the Range-Doppler maps obtained using the FFT and the AR-MEM. Figure \ref{fig:compare_psd} shows such maps along the central bearing for short samples duration ($N=128$ time steps, $T=33~\textrm{s}$). The AR-MEM was computed with the optimal order ($p=64$) and the PSD was evaluated on a very fine grid of 4096 bins, corresponding to a frequency step $\Delta f = 9,4 \cdot 10^{-4}~\textrm{Hz}$. We recall that the analytical expression (eq. \ref{eq:psd_ar}) of the PSD allows for an arbitrary small sampling rate once the AR coefficients have been derived. This is significantly better than the frequency resolution linked to the FFT algorithm which is related to the number of time steps (here, $\Delta f = 2,3 \cdot 10^{-2}~\textrm{Hz}$). Even though improving the sampling rate does not necessary mean improving the resolution, it is clear from the insets of Figure \ref{fig:compare_psd} that a notable improvement of the latter has been reached with the AR-MEM PSD, together with a significant increase of SNR.

\begin{figure}[h!]
    \centering
    \includegraphics[width=\linewidth]{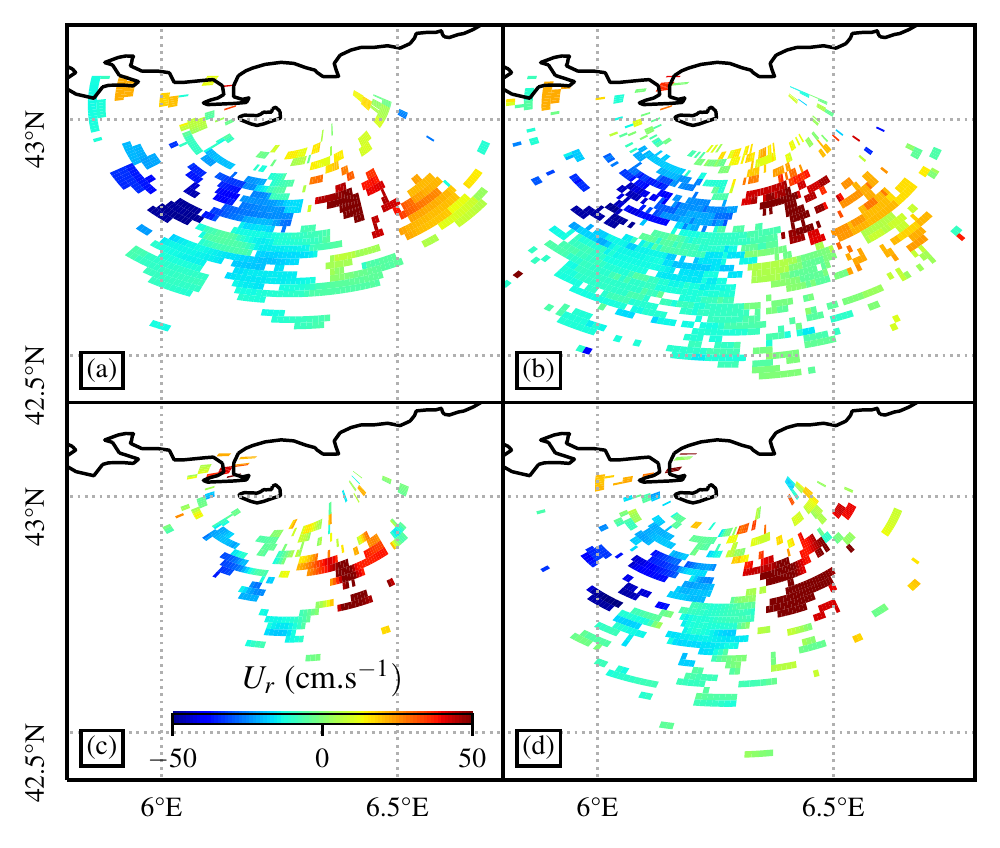}
    \caption{Maps of radial surface current $U_r$ (colorscale; cm.s$^{-1}$) acquired on Feb. 4th, 2019, with the Cap B\'enat WERA HF radar (around 14:01~UTC) and estimated using the: (a)~FFT, samples of $N=4096$ time steps; (b) AR-MEM, samples of $N=4096$ time steps; (c) FFT, samples of $N=128$ time steps; (d) AR-MEM, samples of $N=128$ time steps.}
    \label{fig:carto_comparaisons}
\end{figure}

Figure \ref{fig:carto_comparaisons} shows four radial surface currents maps processed from the same dataset with either the FFT method or AR-MEM. The same QC (eq. \ref{QC}) was used and different sample durations were tested (18 and 0.5~min, respectively). As seen, both methods provide consistent maps but AR-MEM allows for a significant improvement of spatial coverage and range of available radial currents. This amelioration is drastic for short samples where the FFT method gives very sparse results.

The gain in coverage with AR-MEM has been quantified in Table \ref{tab:bistatic_mapping} for different sample durations (18, 4.5 and 0.5~min, respectively), where the number of filled cells has been compared with the maximum expected coverage with a steering beam of 100$\degree$ and a 25~km range (that is about 1500 radar cells at 1$\degree \times$\,1.5~km resolution; see Figure \ref{fig:intro_carto}). As already noted, the coverage is strongly increased, especially for short observation times, from 19 to 45~\% for 0.5 min integration time.

\begin{table}[h]
    \centering
    \caption{Coverage of the sea surface current maps of Figure \ref{fig:carto_comparaisons}}
    \begin{tabular}{lccc}
    \hline\hline\\[-2mm]
    \multirow{2}{*}{\textit{Method}} & \multicolumn{3}{c}{Coverage} \\[1pt]
                            & $N=4096$ & $N=1024$ & $N=128$ \\[1mm] \hline\\[-2mm]
    FFT                     & 75 \% & 59 \% & 19 \% \\
    AR-MEM                    & 85 \% & 78 \% & 45 \% \\[2pt] \hline\hline
    \end{tabular}\\[2mm]
    \begin{minipage}[c]{8cm}
    	\centering\footnotesize{\textit{Note:} Coverage expressed in \% of a standard coverage area, obtained with AR-MEM and FFT, for various integration times ($N=\{4096;1024;128\}$~time steps, i.e. $T\approx\{18;4.5;0.5\}$~min).}
    \end{minipage}
    \label{tab:bistatic_mapping}
\end{table}

\begin{figure}[h]
    \centering
    \includegraphics[width=\linewidth]{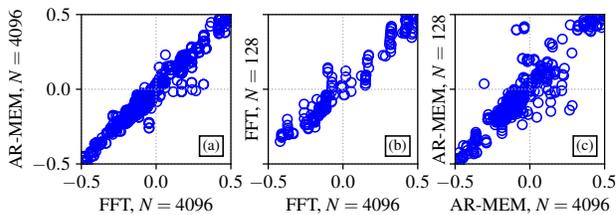}
    \caption{Scatter plots of estimated radial surface currents over the whole covered area on Feb. 4th, 2019 (see Figure \ref{fig:carto_comparaisons}): (a) FFT, long sample (4096 time steps) vs AR-MEM (4096 time steps); (b) AR-MEM, long sample (4096 time steps) vs AR-MEM, short sample (128 time steps); (c) FFT, long sample (4096 time steps) vs FFT, short sample (128 time steps).}
    \label{fig:carto_regressions}
\end{figure}

Together with the gain in coverage it is important to assess the accuracy of the estimated radial currents, especially for short data samples. Deriving an absolute accuracy with respect to some reference measurement (e.g. drifters) is a delicate task because of the difference of scales that enter in play. However, consistency tests can be performed between radial currents estimated with long and short integration times. If one assumes the current to be stationary over the time interval, long integration times are known to be more reliable because of an increased SNR and augmented Doppler frequency resolution. Using the same data set as depicted on Figure \ref{fig:carto_comparaisons}, we performed various estimations of radial currents according to both methods for different integration times and using the same QC (eq. \ref{QC}) as previously. We first checked that the AR-MEM and the FFT method are perfectly consistent for long integration times (4096 time steps or 18~min), as shown on the left panel of Figure \ref{fig:carto_regressions}. Second, short-time estimations were compared to long-time estimations for each method (middle and right panels of Figure \ref{fig:carto_regressions}). As seen, short and long samples lead to very close performances for the AR-MEM while the data passing the QC with the short-time FFT are much sparser.

\begin{figure}[h]
    \centering
    \includegraphics[width=\linewidth]{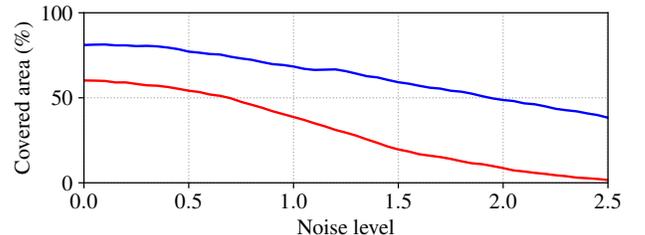}
    \caption{Fraction of radar cells (covered area, in $\%$) passing the quality test when the radial current is computed with short (4.5~min, 1024 time steps) noisy HFR times series and processed with: {\bfseries\color{blue} ------}~AR-MEM; {\bfseries\color{red} ------}~FFT.}
    \label{fig:carto_robustesse}
\end{figure}

Noise robustness was previously assessed using synthetic data and level-varying simulated noise. We tested in a similar way the robustness of real data to the noise level. To do so, we augmented artificially the level of noise by adding a scaled version of the experimental noise obtained with the RFI files. Hence, the actual measured voltage $s(t)$ was transformed to $s(t)+\alpha n(t)$ with unit variance normalization, in a similar way to the synthetic signal in (eq. \ref{eq:signal_bruite}). Again, the covered area is defined as the proportion of radar cells passing the QC, in percent of a maximal expected coverage; it was systematically calculated with the two methods for different levels $\alpha$ of additional noise. The results, shown in Figure \ref{fig:carto_robustesse}, confirm for the two types of noise the significant increase in coverage (or success rate) that was first observed with synthetic data. Note that the coverage is smaller than in the former case for a same value of $\alpha$; in particular, a full coverage is no longer obtained for $\alpha=0$. This is explained by the fact that the experimental data are already noisy before being artificially corrupted by an additional noise. Hence, the actual signal with $\alpha=0$ in fact corresponds to the synthetic case with some $\alpha>0$.

Hence, the AR-MEM allows for a considerable improvement of current mapping both in terms of available coverage and robustness to noise. When discussing the performances of the AR-MEM it is also important to estimate the computational time to calculate the PSD. The classical estimation method based on the FFT algorithm requires $O(N\log\,N)$ operations for a sample of $N$ time steps. When computing the PSD with AR-MEM, the most time-consuming step is the AR parameters estimation since the computation of the PSD (eq. \ref{eq:psd_ar}) also relies on FFT once the AR coefficients have been derived. This operation is known to require $O(Np)$ processing steps \cite{farina1987}, where $p$ is the AR order (eq. \ref{AR}). With $p\sim N/2$ it follows that the complexity of AR-MEM is $O(N^2)$. However, given the short sample size $N$, the computational time remains small; with a standard desktop computer, the time needed to produce the map of Figure \ref{fig:carto_comparaisons}d is of the order of 10~s. For longer integration times such as those typically employed in classical FFT processing, the computational time is significantly increased (about 20~min for $N=4096$, Figure \ref{fig:carto_comparaisons}b) but remains in the reach of real-time processing, all the more that it could be greatly reduced using more adapted hardware. Note that the very interest of using AR-MEM for evaluating the surface current lies in the use of short integration times thus allowing for fast calculations.

\section{Comparisons with surface drifters}\label{sec:drifters}

\begin{figure}[h]
	\centering
	\includegraphics[width=\linewidth]{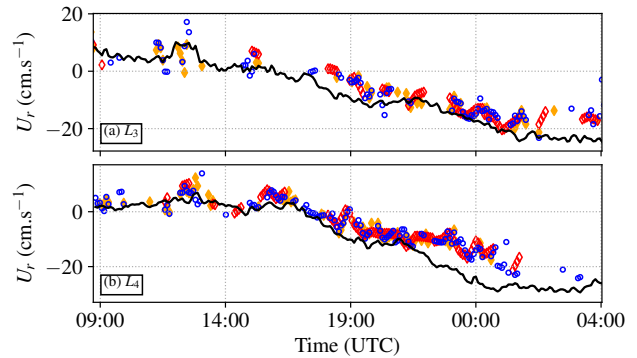}
	\caption{Temporal fluctuations of the radial surface current $U_r$ on June 24-25th, 2019, for the drifters $L_3$ and $L_4$ ({\bfseries\color{black} ------}) and for the Cap B\'enat WERA HFR, estimated with: {\color{blue}{$\circ$}}~AR-MEM and a short integration time of $T=6$~min (1364 time steps); {\color{orange}{$\blacklozenge$}}~FFT and a short integration time; {\color{red}{$\lozenge$}}~FFT and a long integration time of $T=18$~min (4096 time steps).}
	\label{compaflotteurs}
\end{figure}

To assess the performances of the AR-MEM method in operational conditions we performed a comparison of the radial surface currents inferred from HF radar measurements and those measured by Carthe drifters during a recent dedicated campaign in the aera of Toulon. The experiment was conducted on June 24-25th, 2019 and is fully documented in \cite{dumas_OD2020}, where a successful comparison was established with high azimuthal resolution HF radar currents obtained with an improved Direction Finding method \cite{dumas_arxiv2020}. Here, we reprocessed the corresponding radar data with a Beam-Forming technique combined with the AR-MEM method to evaluate the bistatic radial current (transmitter in Porquerolles and receiver in Cap  B\'enat) at a high temporal rate. An integration time of 6 min (one third of standard WERA files) was chosen, in agreement with the 5 min acquisition rate of drifters. The conventional FFT processing with short (6 min) and long (18 min) integration time was also carried out for reference. The results of the different processings is shown in Figure \ref{compaflotteurs} together with the radial currents obtained by an appropriate projection of drifters current vector. We selected the longest 2 drifter trajectories $L_3$ and $L_4$, which can be seen on the Figure 11 from reference \cite{dumas_OD2020} and are now plotted as a function of time. The performances of the different methods when compared with the reference drifters' measurements are recap in Table \ref{tab:perfL3L4}. All three processing are found accurate within 1-2~cm.s$^{-1}$ Root Mean Square Error (RMSE). However, the use of AR-MEM at short integration time (6~min) results in a drastical increase in QC sucess rate (eq. \ref{QC}) with respect to the conventional short-time FFT (from 9-12 \% to 23-25 \%). Note that, despite a comparable success rate to the long-time FFT (18~min), the short-time AR-MEM provides about 3 times more estimations because of its shorter sampling duration.

\begin{table}[h]
    \centering
    \caption{Comparison with drifters}
    \begin{tabular}{lcccc}
    \hline\hline\\[-2mm]
    \multirow{3}{*}{\textit{Method}} & \multicolumn{2}{c}{RMSE}& \multicolumn{2}{c}{Success rate (\%)} \\[1pt]
                            & $L_3$ & $L_4$ & $L_3$ & $L_4$  \\[1mm] \hline\\[-2mm]
    {\color{blue}{$\circ$}}~AR-MEM short (6 min)           &  1.1 &1.7  & 23  & 25  \\
    {\color{orange}{$\blacklozenge$}}~FFT short  (6 min)          &  1.1&  1.7 & 9 & 12  \\
    {\color{red}{$\lozenge$}}~FFT long (18 min)        &  1.1 & 1.9  & 21  & 24  \\[2pt] \hline\hline
    \end{tabular}\\[2mm]
    	\centering\footnotesize{\textit{Note:} RMSE (in cm.s$^{-1}$) between the HFR radial surface current and the drifters' measurements $L_3$ and $L_4$ (Figure \ref{compaflotteurs}), together with the available coverage along the trajectories and the success rate for each method.}
    \label{tab:perfL3L4}
    \vspace{-3mm} 
\end{table}

\section{Temporal fluctuations of surface current}\label{sec:time}

Another striking application of the AR-MEM is the monitoring of sea surface current at short time-scale. We used the same data set, that is the HFR signal measured in Cap B\'enat, to study the complex time series obtained after beamforming for some specific radar cells. In the following experiment, some radar cells located in the central part of the radar coverage were chosen. Continuous sequences were acquired over 27 hours and split into contiguous non-overlapping sub-sequences of either 4096 (about 18 minutes) or 512 (about 2 minutes) time steps corresponding to ``long'' and ``short'' integration time, respectively. These sub-sequences were processed with the AR-MEM and FFT method and the corresponding time series of radial current were produced. Again, a SNR threshold of 12~dB was used to select the Bragg peaks and the same quality criterion (eq. \ref{QC}) was used to retain estimates obtained with either method. In addition, a comparison was also made with recent parametric methods for surface current estimation based on a Maximum Likelihood Estimation (MLE) \cite{guerin2018}. This method has been proven to be useful in the context of rapidly varying strong currents \cite{guerin_OD18} but does not allow for the simple quality check which is used in eq. \ref{QC}. 

\begin{figure*}
    \centering
    \includegraphics[width=\textwidth]{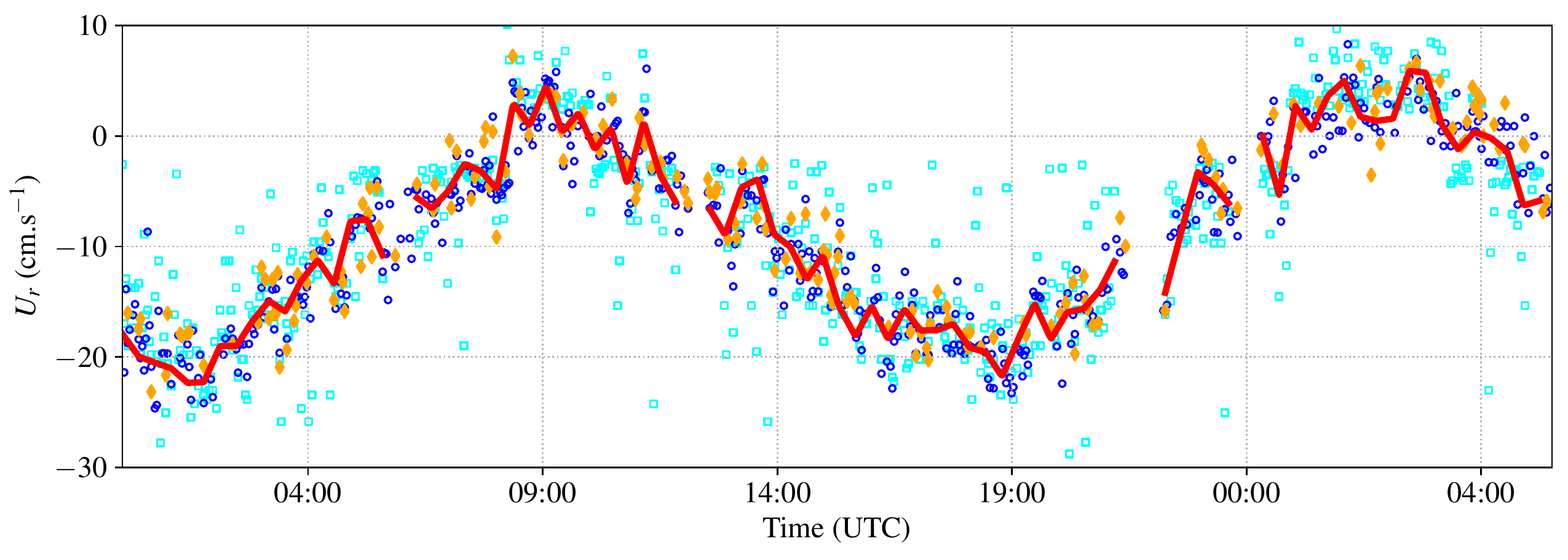}
        \caption{Temporal fluctuations of the radial surface current $U_r$ estimated with a short integration time ($T=132$~s, 512 time steps) according to the: {\color{blue}{$\circ$}}~AR-MEM; {\color{cyan}{$\square$}}~MLE;  {\color{orange}{$\blacklozenge$}}~FFT. The FFT estimate with a long integration time of $T=1065$~s (4096 time steps) is shown for refrence and interpolated over the short time ({\bfseries\color{red} ------}). The data have been acquired on Feb. 4th, 2019 with the Cap B\'enat WERA HFR and the radar cell is chosen in the middle of the coverage.} 
    \label{fig:timeseries}
\end{figure*}

An example of the radial current time series obtained  with different methods and different integration times is shown in Figure \ref{fig:timeseries} for a radar cell located at a range 35~km and at a central azimuth (180$\degree$ from North). As anticipated, the FFT method with short integration time resolves only a few data sequences and therefore does not provide a satisfactory tracking of the sea surface current. When applied to the same short sequences, the AR-MEM successfully resolves the rapid variations of surface current while remaining consistent with the values estimated with a longer integration time (18~min) which can be taken as reference, once linearly interpolated on the short time. The dominant slow oscillation of the measured radial current is consistent with the period of inertial oscillations in the Northwestern Mediterranean sea (about 17 hours) with an amplitude of about 20~cm.s$^{-1}$. Contrarily to the MLE estimate, the fast variations observed with the short-term AR-MEM do not appear to be stochastic and resolve fluctuations of the surface current at the minute time scale. The performances of the different methods when compared to the long-time reference FFT are recap in Table \ref{tab:perftempo}.

\begin{table}[h]
    \centering
    \caption{Comparison with long-time FFT}
    \begin{tabular}{lcccc}
    \hline\hline\\[-2mm]
    \textit{Method (132 s)} & RMSE (cm.s$^{-1}$) & Success rate (\%)  \\[1mm] \hline\\[-2mm]
    {\color{blue}{$\circ$}}~AR-MEM                  &  1.11              & 61 \\
    {\color{orange}{$\blacklozenge$}}~FFT                     &  1.18              & 25  \\
    {\color{cyan}{$\square$}}~MLE                     &  2.32              & 79    \\[2pt] \hline\hline
    \end{tabular}\\[2mm]
    	\centering\footnotesize{\textit{Note:} RMSE (in cm.s$^{-1}$) between the HFR radial surface current estimated with a short integration time ($T=132$~s, 512 time steps) and the reference FFT estimate with a long integration time ($T=1065$~s, 4096 time steps), together with the estimation success rate for each method.}
    \label{tab:perftempo}
\end{table}
 
\begin{figure}[h]
    \centering
    \includegraphics[width=\linewidth]{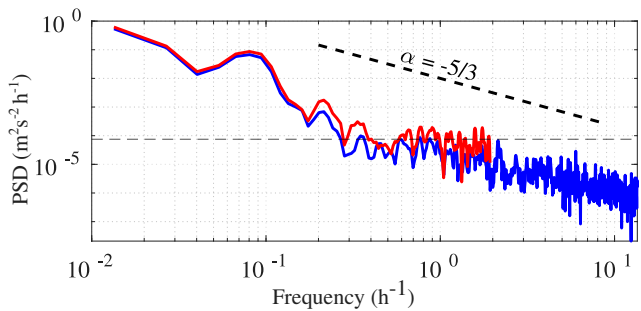}
    \caption{Experimental spectrum of the radial surface current $U_r$ at a given radar cell. $U_r$ estimated with: {\bfseries\color{blue} ------}~MEM, short integration time ($N=1024$~time steps); {\bfseries\color{red} ------}~FFT, long integration time ($N=4096$~time steps). Data have been acquired on Feb. 6th to 10th, 2019, with the Cap B\'enat WERA HFR; the radar cell is in the middle of the coverage.}
    \label{fig:ocean_spectrum}
\end{figure}

As reliable radial surface current estimates can be obtained with a smaller integration time, the corresponding time series can be acquired at a higher temporal frequency compared to the classical FFT estimation. This makes it possible to investigate the PSD of radial surface currents over a wider range of temporal frequencies. Indeed, when reducing the minimal integration time from about 18 to 4.5 minutes, the Nyquist frequency is increased from about 1.7 to 6.7 per hour. The PSD of the radial surface current $U_r$ obtained with the AR-MEM and the FFT method for the same central radial cell as in Figure \ref{fig:timeseries} is shown in Figure \ref{fig:ocean_spectrum} in logarithmic scales. The PSD obtained with the FFT method saturates around the Nyquist frequency and reaches a plateau (thin dashed line) while it pursues its fall off over one extra order of magnitude when estimated with the AR-MEM. The $-5/3$ slope is marked in thick dashed line and is seen to drive the dynamics of the PSD over at least two decades. Such power law decay has already been reported in the literature  \cite{thorpe2007}, 
\begin{equation}
    P(\omega) \propto \omega^\gamma
\end{equation}
where the exponent $\gamma$ depends on the flow regime. The value $\gamma=-5/3$ is typical of Kolmogorov theory of turbulence
and is recovered when inertial forces dominates over viscous.

\section{Discussion and Conclusion}\label{sec:discussion}

We have presented an application of the AR modeling with Maximum Entropy Method (MEM) to HFR mapping and tracking of instantaneous sea surface currents. Contrarily to the common operational approach, the radar signal processing does not use FFT to compute Doppler spectra. Instead, it is based on an AR representation of the complex antenna voltage time series and the calculation of the corresponding AR coefficients with the MEM. As seen, the AR-MEM approach makes it possible to obtain a reliable and robust estimation of the PSD even for short samples. This is particularly suited to the estimation of surface currents which proceeds through the identification of a main resonant frequency in the Doppler spectrum. We have first tested the AR-MEM approach with noisy synthetic HFR data and quantified its augmented performances with respect to the classical spectral estimation in terms of increased coverage and robustness to noise, in particular for short samples.
An application to experimental data has been made with a bistatic HFR located in the region of Toulon. Surface current mapping has been performed with the AR-MEM and FFT method for different integration times. A significant increase of the coverage has been obtained with the former method, in the sense that a larger proportion of current estimates pass the required quality tests. Another useful application has been given, namely the characterization of fast temporal fluctuations of the radial current within a given radar cell. With higher temporal resolution, the power spectrum of surface current fluctuations can be investigated over a wider range of temporal frequencies. The actual high-frequency cut-off can thus be pushed to values as high as a few per minutes and the typical power law decay of turbulence is seen to extend over a few more decades. The ability to estimate surface current variation with an integration time of the order of one minute paves the way to other applications such as tsunami early warning where very short integration times are needed. This will be developed in subsequent work.

\section*{Acknowledgments}
\addcontentsline{toc}{section}{Acknowledgments}

We thank the Direction G\'en\'erale de l'Armement (DGA) and the Agence pour l'Innovation de D\'efense (AID) for funding the Ph.D. grant of B. Domps. The upgrade and maintenance of the WERA HFR system in Toulon have been funded by the EU Interreg program SICOMAR-PLUS and the national observation program MOOSE. We thank the Parc National de Port-Cros (PNPC) for its support and hosting of our radar emitter in Porquerolles Island. We also thank the ``Association Syndicale des Propri\'etaires du Cap B\'enat'' as well as the ``Group Military Conservation'' and the Marine Nationale for hosting our radar installations. Many thanks go to Pr Anne Molcard for providing the drifters' data and for helping with comparisons.


%
\end{document}